\documentclass{iopart}                
 \expandafter\let\csname equation*\endcsname\relax
  \expandafter\let\csname endequation*\endcsname\relax
\usepackage{graphicx,amsmath,amssymb,epstopdf,iopams,cite}
\renewcommand{\v}[1]{{\bf #1}}

\def\be{\begin{equation}}
\def\ee{\end{equation}}

\newcommand{\Eq}[1]{Eq.~(\ref{#1})}

\renewcommand{\Im}{{\rm Im}}

\def\RVB19{3280}
\begin{document}
\title{The impossibility of exactly flat non-trivial Chern bands in strictly local  periodic tight binding models}

\author{Li Chen$^1$, Tahereh Mazaheri$^1$, Alexander Seidel$^1$ and Xiang Tang$^2$}

\address{$^1$  {
Department of Physics, 
Washington University, St. Louis, MO 63130, USA}}

\address{$^2$ {
Department of Mathematics, 
Washington University, St. Louis, MO 63130, USA}}

\begin{abstract}
We investigate the possibility of exactly flat non-trivial Chern bands in tight binding models
with local (strictly short-ranged) hopping parameters.
We demonstrate that while any two of three criteria can be simultaneously realized
(exactly flat band, non-zero Chern number, local hopping), it is not possible
to simultaneously satisfy all three. Our theorem covers 
both the case of a single flat band,
for which we give a rather elementary proof, as well as the case of multiple degenerate
flat bands. In the latter case, our result is obtained as an application of $K$-theory.
We also introduce a class of models on the Lieb lattice with nearest and next-nearest neighbor
hopping parameters, which have an isolated exactly flat band of zero Chern number but, in general, non-zero Berry curvature. 
\end{abstract}

%
%

\section{Introduction} 

The possibility of fractional Chern insulators has been very actively explored recently in models
with  energetically flat bands of non-zero Chern number, and in some cases with consideration of a specific  
particle-particle interaction \cite{Tang11,Sun11,Neu11,Sheng11,  Qi11, Wang11, Regnault11,Ran11,Bernevig12,WangPRL12,WangPRB12,Wu12,Laeuchli12,Bergholtz12,Sarma12,Liu13,bergholtz_review}. 
The non-trivial Chern number makes these models attractive due to
the presence of a quantized Hall effect\cite{TKNN,haldane} and gapless chiral edge modes.
The flatness of the band further increases the resemblance between the physics of interacting
electrons occupying such bands and those occupying a Landau level in the fractional quantum Hall 
regime. More generally speaking, the flatness of a partially occupied band renders the effect of any interactions
non-perturbative, and increases the likelihood of the emergence of interesting correlation effects.
In practice, flatness is always achieved either approximately by tuning hopping parameters in a simple tight binding model \cite{Tang11,Sun11,Neu11,Ran11,Bergholtz12,Sarma12},
or by brute force projection onto a 
given non-trivial Chern band.
As we will review below, the latter case corresponds to passing to hopping parameters that decay exponentially
with distance\cite{Neu11,mcgreevy} (but are not strictly local).

The most desirable playground for the physics of a fractional Chern insulator might thus be given by a 
model featuring an exactly flat non-trivial Chern band arising from a simple, local-hopping tight binding model.
By ``local'', we mean that the range of hopping has an upper bound. Such locality is attractive both from the point of
view of greatest possible simplicity, and also because it may benefit convergence to the thermodynamic limit
in numerical simulations.
Indeed, any combination of two of the three criteria of zero bandwidth, non-trivial Chern number,
and local hopping, can be simultaneously satisfied, as we will review below.
 This might raise the hope that it is
possible to have all three of them satisfied at once.
Previous work has given a qualitative
discussion of flat Chern bands in 
models with {\em exponentially decaying} hopping,
and pointed out that whether truly local
hopping terms may achieve the same is an open question\cite{mcgreevy}.
Other work has conjectured that this may not be possible\cite{Qi}.
Here we present a rigorous proof that the answer to this question is indeed negative.
To be precise, we prove the following 

{\bf Theorem}: If a tight binding model on a periodic two-dimensional lattice with strictly local 
hopping matrix elements has a group of $d$ degenerate flat bands, then the Chern number
associated to this group of bands is zero.

By ``periodic'', we mean that there exist two commuting ordinary lattice translations $T_1$ and $T_2$
generating a two-dimensional lattice group of symmetry operations that leave the Hamiltonian invariant.
We note that while tight binding models describing a periodic magnetic flux through plaquettes of the
lattice are in general not of this form, they can be brought into this form by a gauge transformation
if there exists a ``magnetic unit cell''
through which the net flux is integer.  To such models 
our theorem will also apply. Examples are given, e.g., in Ref. \cite{katsura}.
There, moreover, ``topological flat bands''  have been
constructed  in magnetic systems that do, however, require specific periodic boundary conditions rendering the system a cylinder
or tube with one finite direction, and would not be flat in the two-dimensional thermodynamic limit.
It is always with the two-dimensional thermodynamic limit in mind that we will use the term ``flat'' here.
Topological bands that are flat in the latter sense are known to be possible in magnetic lattice systems
with non-local but super-exponentially decaying hopping matrix elements \cite{kapit, atakisi}.
We further remark that a wealth of three dimensional tight binding models is known that are energetically flat only
along certain lower dimensional cuts \cite{nussinov}. When a local tight binding model is flat within a
two-dimensional cut through the Brillouin zone, obvious generalizations of our theorem will apply.

The remainder of this paper is organized as follows.
In Sec. \ref{prelim} we derive key analytic properties
of the projection operator onto flat bands arising
in local tight binding models. In Sec. \ref{examples}
we discuss some examples of interest, which demonstrate
the analytic features derived in Sec. \ref{prelim}
and help to motivate and anticipate our main result.
In Sec. \ref{nondeg} we give a rather elementary proof of our theorem
for the special case of $d=1$, i.e., non-degenerate flat bands.
In Sec. \ref{deg} we generalize our result to multiply
degenerate flat bands by making use of $K$-theory.
We conclude in Sec. \ref{conclusion}.

\section{Preliminaries\label{prelim}}

We consider two-dimensional periodic tight binding models with $N$ orbitals per unit cell.
The creation operator of a Bloch state in the $n$th band can be written
as 
\be
   c^\dagger_n(\v k) = \sum _{{\v R},i} e^{i{\v k}\cdot {\v R}} u_n^i({\v k}) c^\dagger_{{\v R},i}\,,
\ee
where $c^\dagger_{{\v R},i}$ creates a particle in the orbital labeled $i$ associated with the
unit cell labeled by the lattice vector $\v R$. The coefficients $u_n^i({\v k})$ are obtained as the eigenvectors
of a Hermitian matrix $H_{ij}({\v k})$, and the corresponding eigenvalues $E_n({\v k})$ define the
dispersions  of the $N$ bands of the model. The matrix elements $H_{ij}({\v k})$ are continuous functions
on the torus $T^2$ with which we identify the Brillouin zone,  defined by the $\v k$-plane modulo the reciprocal lattice. 
Indeed, if the model is defined by hopping amplitudes $t_{ij}(\v R_1-\v R_2)$ describing the hopping
from the orbital $(\v R_2,j)$ to the orbital $(\v R_1,i)$, the matrix $H_{ij}(\v k)$ is just obtained as the
Fourier transform
\begin{equation}\label{FT}
     H_{ij}(\v k) = \sum_{\v R} t_{ij}(\v R) e^{-i\v k \cdot \v R}\,.
\end{equation}
These matrix elements will therefore have nice analytic properties whose details depend on the model's hopping parameters
and are evident from \Eq{FT}.
Specifically, if the hopping parameters have a finite range, the $H_{ij}({\v k})$ may be expressed
in (positive and negative) powers of $x\equiv e^{i{\v k}\cdot {\v R_1}}$ and $y\equiv e^{i{\v k}\cdot {\v R_2}}$, with exponents limited by the hopping range.
Here $\v R_1$ and $\v R_2$ are lattice vectors spanning the Bravais lattice underlying the two-dimensional crystal. 
The matrix elements $H_{ij}({\v k})$ are thus given by {\em Laurent polynomials} $H_{ij}(x,y)$,
which evaluated on the torus $T^2=S^1\times S^1$ give the Hamiltonian at $\v k$, i.e., $H_{ij}({\v k})\equiv H_{ij}(e^{i{\v k}\cdot {\v R_1}},e^{i{\v k}\cdot {\v R_2}})$.
We will distinguish between these functions from $T^2$ to $\mathbb{C}$, and the underlying Laurent polynomials 
$H_{ij}(x,y)$, which we will also denote simply by $H_{ij}$ when no confusion may arise.
The same conventions will be adopted for general Laurent polynomials $P\equiv P(x,y)$ and associated functions
  $P({\v k})\equiv P(e^{i{\v k}\cdot {\v R_1}},e^{i{\v k}\cdot {\v R_2}})$.
It will be useful to note, however, that if the functions $P(\v k)$ and $Q(\v k)$ derived from Laurent polynomials
$P$ and $Q$ agree on $T^2$, then  $P$ and $Q$ are already identical as Laurent polynomials. This follows since
$P$ is by definition of the form
\be\label{general_Laurent}
  P= \sum_{m=m_{\sf min}}^{m_{\sf max}}\sum_{n=n_{\sf min}}^{n_{\sf max}} a_{mn}x^m y^n 
\ee
and all the coefficients $a_{mn}$ can be determined by knowledge of $P(\v k)$ on $T^2$, and are 
given by Fourier integrals.

We will now assume that $H_{ij}(\v k)$ describes a model with $d$ isolated 
flat bands, whose energy may be taken to be zero. That is, we have $E_n(\v k)\equiv 0$ for $n=1\dotsc d$, while $E_n(\v k)\neq 0$ 
for all $\v k$ for $n=d+1\dotsc N$. Since the torus is compact, we thus have
$|E_n(\v k)|\geq E_g$ for all $\v k$ and $n>d$, where $E_g>0$ is the gap separating
the flat bands from the remaining ones. We will now describe the projection $P_{ij}(\v k)$ onto the zero energy (flat band) subspace at $\v k$,
and find that it is essentially also of Laurent polynomial form, up to an overall normalization function. 
To this end, we write
\be\label{Heps}
 H_\epsilon(\v k) =   H(\v k)+\epsilon 1\!\!1 = \epsilon P(\v k) + \sum_{E\neq 0} (E+\epsilon) P_E(\v k)\,.
\ee
Here, 
$P(\v k)$ is the orthogonal projection onto the $d$-dimensional zero energy subspace, the sum goes over all
non-zero eigenvalues of $H(\v k)$, and $P_E(\v k)$ is the orthogonal projection
onto the eigenspace corresponding to the eigenvalue $E\neq 0$.
For $0<\epsilon<E_g$, $H_\epsilon(\v k)$ is invertible, and it is seen from \Eq{Heps} that
\be\label{Pk0}
   P(\v k) = \lim _{\epsilon\rightarrow 0} \epsilon H_\epsilon^{-1}(\v k) =\lim _{\epsilon\rightarrow 0} \frac{\epsilon}{\det H_\epsilon(\v k) } \text{adj}(H_\epsilon (\v k)) \,,
\ee
where for the second equality, we have written the inverse of $H_\epsilon(\v k)$ in terms of its inverse determinant
and the adjugate matrix of 
$H_\epsilon(\v k)$. By \Eq{Heps}, this limit is well-defined, and so is 
\be\label{fk}
    { f(\v k)}=  \lim _{\epsilon\rightarrow 0}\, \frac {\det H_\epsilon(\v k) }{\epsilon^d} =\prod_{n=d+1}^N E_n(\v k)\,.
\ee 
$f(\v k)$ is clearly a continuous and nowhere vanishing function from the torus $T^2$ (the Brillouin zone) to $\mathbb{R}$.
Because of the existence of the limit \eqref{fk}, we may rewrite \Eq{Pk0} as
\be  \label{Pk}
      P(\v k) = \frac{1}{f(\v k)} \lim_{\epsilon\rightarrow 0} \epsilon^{1-d}  \text{adj}(H_\epsilon (\v k)) = \frac{h(\v k)}{f(\v k)}\,,
\ee 
where together with the limits in Eqs. \eqref{Pk0}, \eqref{fk}, the limit in this last equation must also be well-defined,
since $f(\v k)\neq 0$. We have denoted this matrix-valued limit by $h(\v k)$. We will now show that 
the functions $f(\v k)$ and $h_{ij}(\v k)$ are also of Laurent-polynomial form, just as the original
matrix elements $H_{ij}(\v k)$. We denote the ring of Laurent-Polynomials in $m$ variables
by $\mathbb{C}[x_1^{\pm 1},\dotsc,x_m^{\pm 1}]$. Then clearly $f_\epsilon(k)=\epsilon^{-d} \det H_\epsilon(\v k)$
is obtained from a Laurent-polynomial $f_\epsilon\in \mathbb{C}[x^{\pm 1},y^{\pm 1},\epsilon^{\pm 1}]$ upon substitution
of  $x= e^{i{\v k}\cdot {\v R_1}}$ and $y=e^{i{\v k}\cdot {\v R_2}}$, since the matrix elements of
$H_\epsilon(\v k)$ have this property. Moreover, since the limit in \Eq{fk} is finite everywhere on $T^2$,
no negative powers of $\epsilon$ can appear in $f_\epsilon$. (Indeed, the coefficient 
of $\epsilon^{-n}$ would be an element of $\mathbb{C}[x^{\pm 1},y^{\pm 1}]$, which, in order to vanish
on the entire torus $T^2$, must be the zero polynomial, as we remarked above.)
Thus $f_0$ is well-defined and in $\mathbb{C}[x^{\pm 1},y^{\pm 1}]$, and by \Eq{fk}, $f(\v k)=f_0(\v k)$.
We will also write $f$ instead of $f_0$ from now on. From \Eq{Pk}, we can apply the same reasoning to
the matrix elements of $\epsilon^{1-d}  \text{adj}(H_\epsilon (\v k))$. This leads to the conclusion
that the matrix elements of $h(\v k)$
are likewise given by Laurent-polynomials $h_{ij} \in \mathbb{C}[x^{\pm 1},y^{\pm 1}]$ 
evaluated at $x= e^{i{\v k}\cdot {\v R_1}}$ and $y=e^{i{\v k}\cdot {\v R_2}}$.
In sections \ref{nondeg} and \ref{deg}, we will prove that projection operators of the form \eqref{Pk},
with $h_{ij}(\v k)$, $f(\v k)$ both given by Laurent-polynomials, describe complex
vector bundles over the torus with zero Chern number.
In the following section, we will first give some examples of the projection operators
$P(\v k)$ for some models of interest in the present context.

\section{Examples of models with non-degenerate exactly flat bands\label{examples}}

Before we proceed to investigate the Chern number of flat bands 
arising in finite range hopping models, we 
give some examples for the detailed structure
of the projection operator $P(\v k)$ arising in models
belonging to this and other related classes.
As we asserted in the introduction, for an energy band
in a tight binding model, any two
of the following  three desirable properties
may simultaneously hold:
 non-zero Chern number, flat dispersion, and
 the fact that the underlying model is local.
 Historically, the most important demonstration 
 of such properties was given by the Haldane-model \cite{haldane},
 a two-band model defined on a honeycomb lattice with only
 nearest and next-nearest neighbor hoppings, which has non-trivial Chern bands.
 From here, one easily obtains a new, non-local hopping model
 on the same lattice, featuring the exact same Bloch-type eigenstates
 defining two non-trivial Chern bands, but each having a flat dispersion.
 For this one need only consider the projection operator onto one of the
 two bands, again denoted by $P(\v k)$.
 We consider a Haldane model with hopping parameters $t_1$ and $it_2$ (hence flux $\phi=\pi/2$),
 and on-site parameter $M=0$. 
 From an explicit solution for the eigenstates at $\v k$, one immediately obtains
 \begin{subequations}\label{haldane}
 \be
   \renewcommand{\arraystretch}{2}
   P(\v k)=
 \begin{pmatrix}
    \frac{\varepsilon+\alpha}{2\varepsilon} & \frac{\beta^{\ast}}{2\varepsilon}  \\
  \frac{\beta}{2\varepsilon} & \frac{\varepsilon-\alpha}{2\varepsilon} 
 \end{pmatrix}\,,
 \ee
  where
 \be
   \alpha =2t_{2} \sum _{\ell=1}^3 \sin({{\v k}\cdot {\v R_{\ell}}})\,,
\ee
\be
   \beta =t_{1} (1+e^{i{\v k}\cdot {(\v R_{1}+\v R_{2})}}+e^{i{\v k}\cdot {\v R_{2}}} )\,,
\ee
\be\label{epsilon_haldane}
   \varepsilon = \sqrt{\alpha^2+\beta^\ast\beta} \,,
\ee
\end{subequations}
and, following Haldane \cite{haldane}, we chose primitive lattice vectors $R_1$ and $R_2$ at $120^\circ$, angle, 
and also let $R_3=-R_1-R_2$.
 We may now interpret the above matrix-valued function as the Hamiltonian at $\v k$
 of a different, non-local hopping model, writing $H(\v k)=P(\v k)$.
 This model then has two flat Chern-bands at energies $0$ and $1$, respectively.
 Furthermore, the analytic structure of the matrix elements in \Eq{haldane}
 clearly features branch cuts as a function of $x= e^{i{\v k}\cdot {\v R_1}}$ and $y=e^{i{\v k}\cdot {\v R_2}}$
 when analytically continued to values with $|x|,|y|\neq 1$.
 This  implies that \Eq{haldane} is not of the form \eqref{Pk}, since such branch
 cuts do not arise in rational functions of Laurent-polynomials.
 Thus, the real-space Hamiltonian associated with $H(\v k)=P(\v k)$ is not
 strictly finite ranged. Rather, it is exponentially decaying. This Hamiltonian
 belongs to a class studied in Ref. \cite{Qi}, where lower bounds have been derived on 
 moments of the distribution of hopping parameters.

 The fact that the hopping range of $H(\v k)=P(\v k)$ decays exponentially
 may be seen as follows.
 As an operator acting on the full Hilbert space, the projector onto the upper energy  band
 can be written as
 \begin{equation}\label{haldane2}
\hat H= \hat P= \sum_{\alpha,\beta=0,1} \iint d^2kd^2k' \, \langle\Omega|   c_\alpha (\v k)c_\beta^\dagger(\v k') |\Omega\rangle\,  c_\alpha^\dagger  (\v k)c_\beta(\v k') \,.
 \end{equation}
Here, the operator $c_\alpha^\dagger (\v k)$ creates a Bloch-state in the lower (upper) band of the Haldane model
for $\alpha=0$ ($1$), $|\Omega\rangle$ is the ground state of the Haldane model at half filling, and the integrals are over the Brillouin zone.
The structure of \Eq{haldane2} is clearly basis-invariant. Hence, when we re-express
the operators $c^\dagger_\alpha(\v k)$ through operators creating a particle 
in the orbital labeled $i$ at the unit cell labeled $\v R$, $c^\dagger_{\v R,i}$,
we get
\be
\hat H= \hat P= \sum_{i,j=1,2} \sum_{\v R,\v R'}\, \langle\Omega|   c_{\v R, i} c_{\v R',j}^\dagger |\Omega\rangle\,  c_{\v R,i}^\dagger  c_{\v R',j} \,.
\ee
Now, the hopping matrix element $\langle\Omega|   c_{\v R, i} c_{\v R',j}^\dagger |\Omega\rangle$ is an equal time
correlation function in a gapped ground state of a local Hamiltonian. It must therefore decay exponentially
with $|\v R-\v R'|$ \cite{hastings} .
More directly, we may study this exponential decay as follows.
The coefficients $\langle\Omega|   c_{\v R, i} c_{\v R',j}^\dagger |\Omega\rangle$
are proportional to the Fourier integral $\int d^2k \, P_{ij}(\v k) e^{i{\v k\cdot(\v R-\v R')}}$,
according to \eqref{FT}.
If viewed as analytic functions in the variables $v=\v k\cdot \v R_1$, $w=\v k\cdot \v R_2$,
suppose 
that the matrix elements $P_{ij}$ are holomorphic in $v$ and $w$
within the domain $\{(v,w)\in \mathbb{C}^2: |\Im\, v|<\Delta_1, |\Im \, w|<\Delta_2\}$.
The largest possible values for $\Delta_1$ and $\Delta_2$ are determined from
the zeros in the quantity $\epsilon$ in \Eq{haldane}. Specifically, simple estimates show that
$\Delta_1=\Delta_2= (\sqrt{3}/8)\log[1+{\varepsilon_g^2}/({3\sqrt{2}(12t^{2}_{2}+t^{2}_{1}))}]$
satisfy the above requirement, though certainly, these are not
optimal choices. Here, $\varepsilon_g=\min(3\sqrt{3} t_2,t_1)$ equals the minimum
of the dispersion \eqref{epsilon_haldane} for real $\v k$. 
It then follows immediately from elementary considerations that for any choice of 
$\Delta_1'<\Delta_1$ and $\Delta_2'<\Delta_2$, there is a constant $C>0$
such that
\be\label{bound}
|\langle\Omega|   c_{\v R, i} c_{\v R',j}^\dagger |\Omega\rangle| < Ce^{-|m|\Delta_1'-|n|\Delta_2'}\,,
\ee
where $\v R-\v R'=m \v R_1+ n\v R_2$.
We note that the bound in \Eq{bound} does not have all the symmetries of the honeycomb lattice,
due to the preference given to primitive lattice vectors $\v R_1$, $\v R_2$, and can be improved already
by symmetrization.

We finally discuss a model with a flat band emerging from local hopping matrix elements.
A trivial example can be obtained by adding an additional orbital to every unit
cell in any given tight-binding model, and leaving these additional orbitals decoupled.
This trivially introduces a flat band corresponding to an ``atomic limit'', i.e., with
completely localized Wannier states that are zero energy eigenstates.
However, we prefer to give a less trivial example, where the emerging
flat band does not have Wannier states of compact support and has a Berry curvature which does not vanish identically.

\begin{figure}
\centering
\includegraphics[width=7cm]{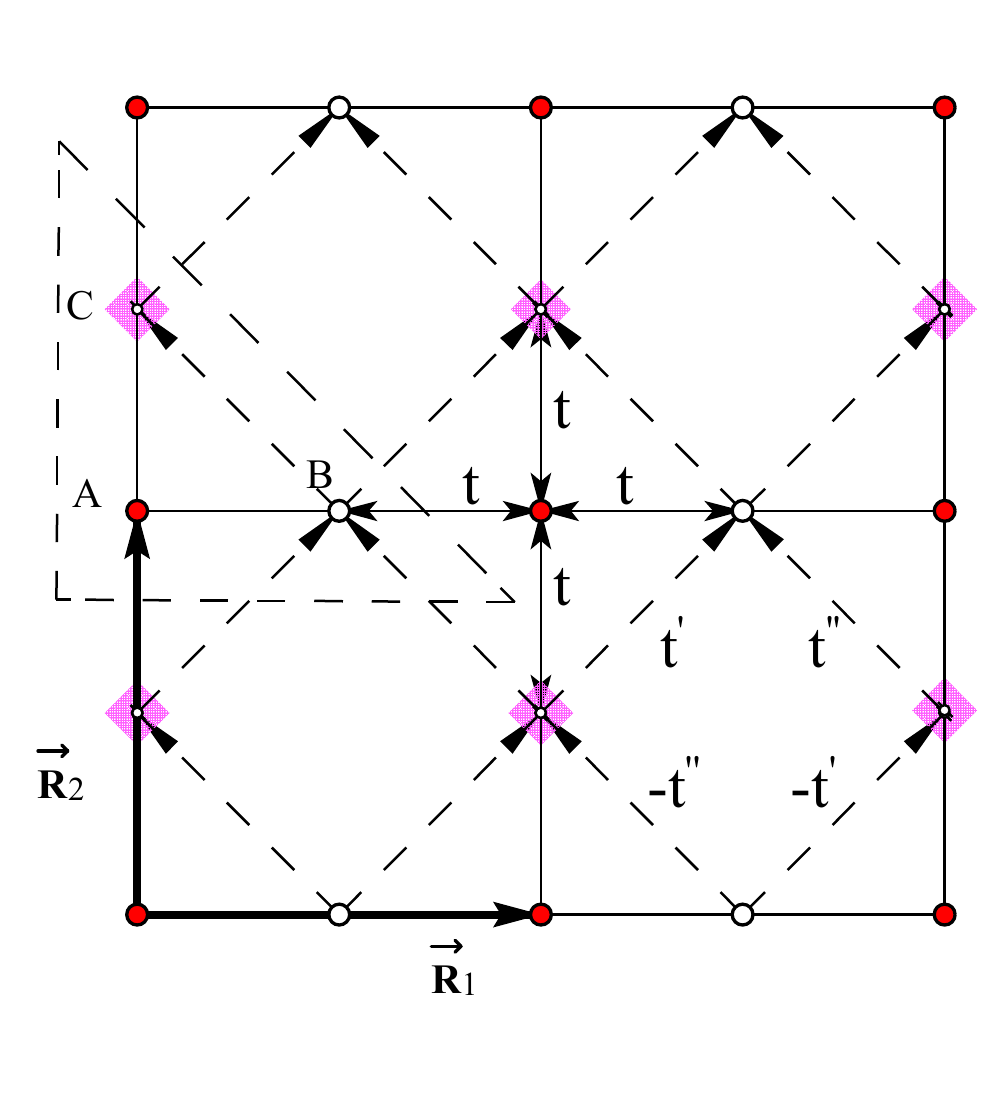}
\caption{A three-band model on the Lieb lattice with three atoms per unit cell. This tight binding model has real nearest neighbor hopping
parameter $t$, and complex next-nearest neighbor hopping parameters $t'$ and $t''$ connecting the sites B and C in the directions specified by  dashed lines.
The model is invariant under a combined particle-hole transformation and inversion with respect to, e.g., plaquette centers.
This guarantees a  zero energy flat band. The latter is separated from the top and bottom band by an energy gap as long as
the two diagonal hopping amplitudes have different imaginary parts.

}
\label{fig1}
\end{figure}

To this end, we consider a three-band model defined on the Lieb lattice,
Fig. \ref{fig1}. The model has real nearest neighbor hopping
amplitude $t$, and two complex next-nearest neighbor
hoppings $t'$ and $t''$ as shown in the figure. 
The Hamiltonian at $\v k$
for this model is given by the matrix
\be
H(\v k)=
 \begin{pmatrix}
  0  & t(1+e^{-ik_x}) &t(1+e^{-ik_y})  \\
  t(1+e^{ik_x}) & 0 & - {{t'}^ * }{e^{i{k_x}}} + t'{e^{ - i{k_y}}} - t''^* + t''{e^{i{k_x} - i{k_y}}}\\
  t(1+e^{ik_y})& - t'{e^{-i{k_x}}} + t'^*{e^{ i{k_y}}} - t'' + t''^*{e^{-{k_x}+i{k_y}}}& 0
 \end{pmatrix}\,,
 \ee 
 where $k_x={\mathbf{k}} \cdot {\mathbf{R_1}}$ and $k_y={\mathbf{k}} \cdot {\mathbf{R_2}}$.
The model has a combined particle-hole/inversion symmetry of the form
\be\label{ph}
   M(\v k)C H(\v k) C M(\v k)^\dagger= -H(\v k)\,,
\ee
where $C$ is the anti-linear operator associated with complex-conjugation,
and $M(\v k)$ is the following matrix:
 \renewcommand{\arraystretch}{1}
\be
  M(\v k)=
 \begin{pmatrix}
  -1 & 0 &0  \\
  0 & e^{ik_x} &0\\
  0& 0& e^{ik_y}
 \end{pmatrix}\,.
 \ee
 \renewcommand{\arraystretch}{1.5}
The symmetry \eqref{ph} implies that if $E_{\v k}$ is an eigenvalue
at given $\v k$, then so is $-E_{\v k}$.
\begin{figure}
\centering
\includegraphics[width=12cm]{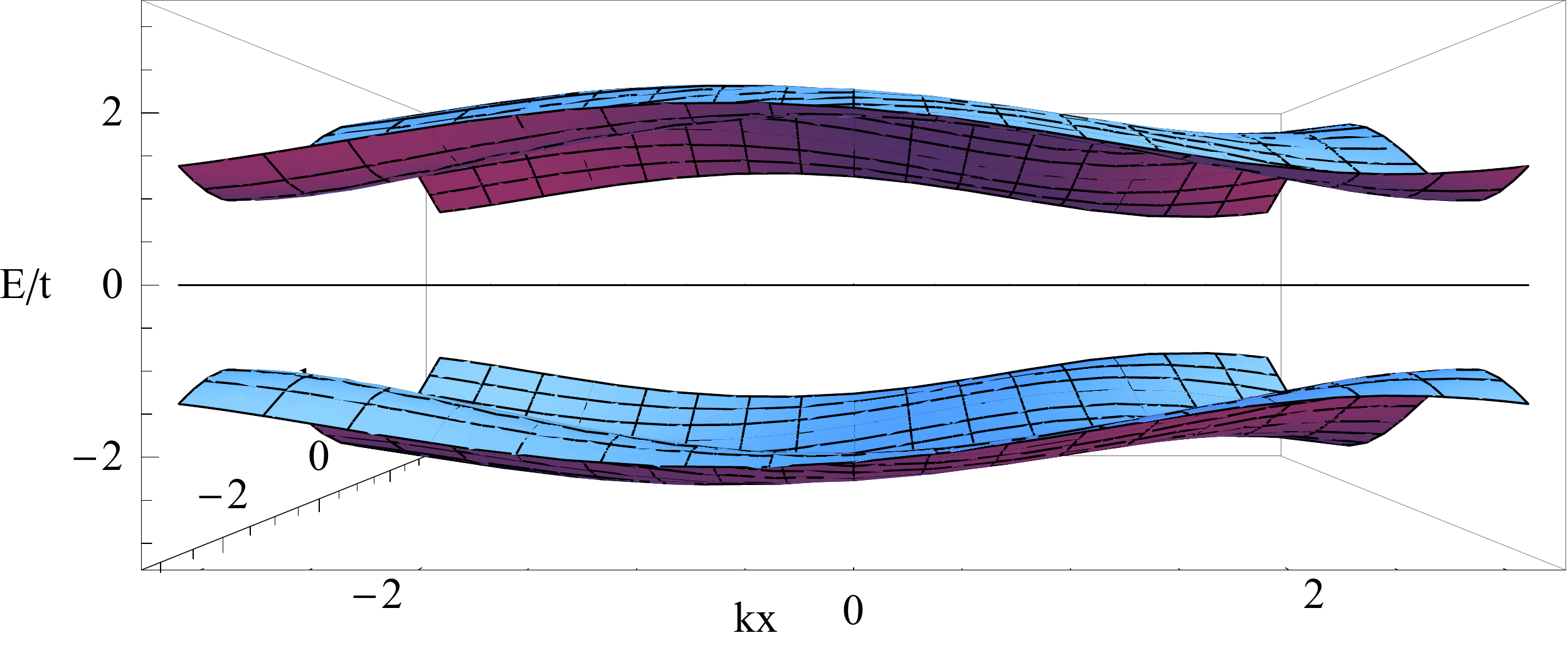}
\caption{The band structure of the Lieb lattice model with $u=0.2$, $v=0.6$, $w=-0.5$ and $s=-0.1$. The energy of the middle band is exactly zero while the upper and lower band have opposite energies.
The Chern number of the flat band vanishes in accordance with our general results.}
\label{fig2}
\end{figure}
When the total number of bands is odd, as in the present case, this necessitates the existence of
a zero eigenvalue at all $\v k$, hence a zero energy flat band.
For the present model, we find that this flat band is isolated by a gap from the other
two bands for any 
$t'$ and $t''$ that have unequal imaginary parts. The upper and lower band have energies, in units of $t$, of 
$\varepsilon_\pm= \pm \sqrt 2 (2sw\sin \left( {{k_x} - {k_y}} \right) + \left( {{s^2} - {w^2}} \right)\cos \left( {{k_x} - {k_y}} \right) + \left( {2vw - 2su} \right)\sin {k_x} + (2sv + 2uw + 1)\cos {k_x}- 2\left( {su + vw} \right)\sin {k_y} + \left( {2\left( {sv - uw} \right) + 1} \right)\cos {k_y} - 2uv\sin \left( {{k_x} + {k_y}} \right) + \left( {{v^2} - {u^2}} \right)\cos \left( {{k_x} + {k_y}} \right) + {s^2} + {u^2} + {v^2} + {w^2} + 2)^{1/2}$, where $u=\operatorname{Re}(t'/t)$, $v=\operatorname{Im}(t'/t)$, $w=\operatorname{Re}(t''/t)$ and $s=\operatorname{Im}(t''/t)$. A typical band structure is shown in Fig.
\ref{fig2} with $u=0.2$, $v=0.6$, $w=-0.5$ and $s=-0.1$.

The projection operator onto the flat band at $\v k$ takes on the following form:

\begin{equation}
  P(\v k)=
 \frac{1}{\varepsilon_+\varepsilon_-}
\left( {\begin{array}{*{20}{c}}
  {{x^{ - 1}}{y^{ - 1}}{g^2}}&{{x^{ - 1}}{y^{ - 1}}\left( {1 + y} \right)g}&{ - {x^{ - 1}}{y^{ - 1}}\left( {1 + x} \right)g} \\ 
  { - {y^{ - 1}}\left( {1 + y} \right)g}&{ - {y^{ - 1}}{{\left( {1 + y} \right)}^2}}&{{y^{ - 1}}\left( {1 + x} \right)\left( {1 + y} \right)} \\ 
  {{x^{ - 1}}\left( {1 + x} \right)g}&{{x^{ - 1}}\left( {1 + x} \right)\left( {1 + y} \right)}&{ - {x^{ - 1}}{{\left( {1 + x} \right)}^2}} 
\end{array}} \right)\,,
 \end{equation}
 where here we let $x=e^{ik_x}$, $y=e^{ik_y}$ and $g={\left( { - t' - t''x + {{t''}^ * }y + {{t'}^ * }xy} \right)/t}$.
We note that this is precisely of the analytic form that follows from \Eq{Pk}. (This would 
not be true for the projection operators onto the non-flat bands of this model, which have Chern numbers
$\pm 1$, respectively.)
It will follow further from the proof presented in the next section that the Chern number
of this flat band must then be zero. In the present case, 
this already follows
from the invariance of the flat band under an anti-linear operator of the form \eqref{ph}.
We note further that while the ``atomic'' limit of the flat band can be reached by sending $t\rightarrow 0$, 
for $t\neq 0$ the Berry curvature in the present model is non-zero in general,
and moreover, its Wannier states
cannot be expected to have compact support (though there exist non-orthogonal zero energy modes
supported on a single square plaquette).
Thus already nearest neighbor density-density interactions, or,
if we add back spin, on-site interactions would have highly non-trivial
effects, no matter how small the interaction strength, whenever the middle band is at fractional filling.
In addition, if we set $t''=0$, the remaining non-zero hoppings give the lattice the topology
of a kagome lattice, and can then all be regarded as nearest neighbor.
Thus, despite its flat band having zero Chern number, the model
may be quite promising as a minimalistic playground for strong interaction
effects. We leave a detailed analysis for future work.\footnote{In the case of repulsive on-site interactions, we expect general
arguments\cite{katsura} in favor of ferromagnetism to apply.}

\section{Proof for the non-degenerate case\label{nondeg}}

We now turn back our attention to the proof that the analytic structure
displayed in \Eq{Pk} implies the vanishing of the Chern number
of the flat band(s) described by the range of the projection operator $P(\v k)$.
We first focus on the case of a single non-degenerate flat band, since this is usually the case of interest,
motivated by the models of the preceding section for example.
Furthermore, for this important special case 
we can give a rather elementary proof.

In this case, the projection operator $P(\v k)$ of \Eq{Pk} is of rank $1$, and its matrix 
elements may thus be written as 
\be\label{ri}
P_{ij}(\v k)= u_i(\v k) u_j(\v k)^\ast.
\ee
Here, we make at first no assumption about the continuity of the functions $u_i(\v k)$, except that
$P_{ij}(\v k)$ must of course be continuous. 
Clearly, the  $u_i(\v k)$ are defined uniquely only up to an overall $\v k$-dependent phase.
Furthermore, $\Tr P_{ij}(\v k) =1$, hence the component functions
$u_i(\v k)$ cannot all simultaneously vanish at any particular $\v k$.

We will show in the following that, up to a normalization factor which is nowhere vanishing on the 2-torus $T^2$, the $u_i(\v k)$
can be chosen to be of Laurent-polynomial form. 
This implies that the $u_i(\v k)$ are smooth functions having the periodicity of the reciprocal lattice.
They thus define a global section of the bundle, which is normalized and nowhere vanishes on $T^2$.
The existence of such a global section implies that the bundle is trivial and hence its Chern number is zero.
Alternatively,
if $u(\v k)$ is the column vector with components $u_i(\v k)$,
 this Chern number is proportional
to the line integral of $A(\v k)= u(\v k)^\ast  \cdot \nabla_{\v k} u(\v k)$ around the Brillouin zone boundary,
which will vanish due to the periodicity of $u(\v k)$. The latter again follows from the Laurent-polynomial structure 
of $u(\v k)$ which we will now prove.

Since $P(\v k)$ is a rank $1$ projection operator, using \Eq{Pk} we have $\det [f^{-1}(\v k) h(\v k)-1\!\!1]=0$ or
$\det[ h(\v k)-f(\v k) 1\!\!1]=0$ for all $\v k$. The latter is a Laurent-polynomial expression evaluated
on the torus. As remarked initially, it follows from this that already $\det [h-f 1\!\!1]=0$ in $\mathbb{C}[x^{\pm 1},y^{\pm 1}]$.
Then, there must be a nontrivial element $q=(q_1,\dotsc,q_N)^t$ in  $\mathbb{C}[x^{\pm 1},y^{\pm 1}]^N$
such that $[ h-f 1\!\!1] q=0$ \cite{AlgebraI}. This implies that
\be \label{Pviaq}
    P_{ij}(\v k)= \frac{h_{ij}(\v k)}{f(\v k)} = \frac{q_i(\v k)q_j^\ast (\v k)}{||q||^2(\v k)} 
\ee 
for all $\v k$ for which the right hand side is well defined.
Here, we have introduced  $||q||^2=\sum_{i=1}^N q_i^\ast q_i$,
and a ring endomorphism  of 
$\mathbb{C}[x^{\pm 1} ,y^{\pm 1}]$ denoted by ``$\ast$'', defined via
\begin{equation}
    \left( \sum_{m=m_{\sf min}}^{m_{\sf max}}\sum_{n=n_{\sf min}}^{n_{\sf max}} a_{mn}x^m y^n  \right)^\ast
    = \sum_{m=m_{\sf min}}^{m_{\sf max}}\sum_{n=n_{\sf min}}^{n_{\sf max}} a_{mn}^\ast x^{-m} y^{-n}.
\end{equation}
Apparently, $\ast$ intertwines with evaluation on the torus, where for the 
associated functions on $T^2$, $\ast$ is just ordinary complex conjugation.

Naively, one would now like to let $u_i(\v k)= q_i(\v k)/\sqrt{||q||^2(\v k)}$.
However, it is not obvious at this point that this is well defined for all $\v k$.
To establish this, we must show that we may assume $||q||^2(\v k)\neq 0$
for all $\v k$. From \Eq{Pviaq}, we can conclude that
\be
   h_{ij}(\v k) ||q||^2(\v k) = f(\v k) q_i(\v k)q_j^\ast (\v k)
\ee
for {\em all} $\v k$. This again implies
\be\label{division}
 h_{ij} ||q||^2 = f q_iq_j^\ast 
\ee
in $\mathbb{C}[x^{\pm 1},y^{\pm 1}]$. The latter is a unique factorization domain (UFD).
If $||q||^2(\v k)=0$ for some $\v k$, there must be some prime factor
$p$ in the factorization of $||q||^2$ such that $p(\v k)=0$ at this particular $\v k$.
Since $f$ cannot have such prime factors, 
$p$ must divide $q_i q_j^\ast$ for all
$i,j$.  We will show now that, actually, 
we have $q_i q_j^\ast=pp^\ast \tilde q_i \tilde q_j^\ast$,
with  $\tilde q_i ,\tilde q_j^\ast\in \mathbb{C}[x^{\pm 1},y^{\pm 1}]$.
To this end, 
it is useful to distinguish the cases $p\simeq p^\ast$ and
$p \not\simeq p^\ast$, where, by $\simeq$, we mean equality
up to a unit (invertible) element.
Consider  $p\simeq p^\ast$  first.
In this case, if $p\nmid q_i$ for some $i$, then we also have
$p\nmid q_i^\ast$. This gives a contradiction when considering
$i=j$ in \Eq{division}. Hence we have $q_i=p\tilde q_i$ for all $i$.
Now, if $p \not\simeq p^\ast$, then if $p|q_i$ for all
$i$, we are done.
On the other hand, if $p \nmid q_i$ for some $i$,
then by \Eq{division}, $p | q_j^\ast$ for all $j$. So we have
$q_j^\ast= p\tilde q_j^\ast$, or $q_j= p^\ast \tilde q_j$, which
concludes the proof that $q_i q_j^\ast=pp^\ast \tilde q_i \tilde q_j^\ast$
with appropriate $\tilde q_i, \tilde q_j$ in all possible cases.
Thus \Eq{division} can be divided by $pp^\ast$ in $\mathbb{C}[x^{\pm 1},y^{\pm 1}]$, where by construction
$||q||^2=pp^\ast ||\tilde q||^2$. This yields
\be\label{canceled}
    h_{ij} ||\tilde q||^2 = f \tilde q_i\tilde q_j^\ast \,,
\ee
and $||\tilde q||^2$ has fewer powers of $p$ than $||q||^2$ .
By iterating this procedure, we obtain $\tilde q_i$ satisfying
\Eq{canceled}
such that no prime element $p$ that somewhere vanishes on 
$T^2$ divides $||\tilde q||^2$. Thus, $||\tilde q||^2(\v k)$ does not
vanish for any $\v k$.
Then $u_i(\v k)= \tilde q_i(\v k) /\sqrt{||\tilde q(\v k)||^2}$ is well defined
for all $\v k$, satisfies \Eq{ri}, and is a smooth non-vanishing global section
 on $T^2$,
thus showing that the Chern number of the flat band projected on by $P$ 
is zero. 

\section{Proof for the degenerate case\label{deg}}

We now consider the general case of $d\geq 1$ degenerate flat bands.
$P(\v k)$ is now a projection operator of rank $d$, and \Eq{ri} must be generalized to
\be\label{rin}
 P_{ij}(\v k)=\sum_{n=1}^d u_i^n(\v k) u_j^n(\v k)^\ast\,.
\ee
Here, the $d$  vectors $u^n(\v k )=(u^n_1(\v k),\dotsc,u^n_N(\v k))^t$ 
correspond to an orthonormal basis of the flat band subspace at $\v k$.
The sum on the right hand side of \Eq{rin} makes the unique factorization
property of the ring $\mathbb{C}[x^{\pm 1},y^{\pm 1}]$ considerably more difficult to use efficiently
when $d>1$.
For this reason, we are not able to give a straightforward generalization of the
elementary approach taken in the preceding section.
Instead we will appeal now to the fact that the ring $\mathbb{C}[x^{\pm 1},y^{\pm 1}]$,
which we will simply denote as $R$ in this section, is regular Noetherian,
and make contact with results from $K$-theory. For this section, we assume that the
reader is familiar with basic notions of $K$-theory, and refer to Ref. \cite{bass}
for a general introduction.

The $d$-dimensional vector bundle 
associated to our flat bands can be thought of as
$P(\v k) C(T^2)^N$, where $C(T^2)$ denotes the ring of continuous functions on the torus, and
$C(T^2)^N$ the free $C(T^2)$-module of dimension $N$. $P(\v k)$ naturally acts on
$C(T^2)^N$ in a $C(T^2)$-linear fashion.
Thus $P(\v k) C(T^2)^N$ is the projective
module which is the range of $P(\v k)$ in $C(T^2)^N$.
We wish to study the $K$-theory element $[P(\v k) C(T^2)^N]$
corresponding to this projective module in $K_0(C(T^2))$.
To this end, we first consider the algebraically
simpler ring $R$ which is naturally embedded
in $C(T^2)$ through the evaluation map we have frequently utilized above.
We have also remarked already that this embedding is indeed injective.
For the ring $R$ we have a theorem due to Swan \cite{swan},
which generalizes Quillen's proof of Serre's conjecture \cite{quillen},
and which implies that every finitely generated projective $R$-module is free.
Note that it is here where the Laurent-polynomial structure of $R$,
which is characteristic of Hamiltonian matrix elements
in finite range tight binding models, crucially
enters, just as it did in the $d=1$ case studied above.

There is a close connection between the $R$-module
$h R^N$ and $P(\v k) C(T^2)^N$, where elements of the
former can be naturally identified with elements in the latter,
again through the evaluation map. Here, $h\in M_N(R)$
is again the $N\times N$ matrix with entries in $R$ defined by
\Eq{Pk}. However, since $h^2=f h$, and $f$ does not necessarily have an inverse in $R$,
we cannot say that $h R^N$  is projective.
Hence, we must work inside the ring $R[f^{-1}]$,
the localization of $R$ by powers of $f\in R$,
which is ``in between'' $R$ and $C(T^2)$.
The functor $\pi=R[f^{-1}]\otimes_R$ sends $R$-modules
to $R[f^{-1}]$-modules
and induces a group homomorphism between
$K_0(R)$ and $K_0(R[f^{-1}])$.
This homomorphism is surjective.
Indeed, since R is regular Noetherian, Corollary 6.4 of Ref. \cite{bass}
applies, and yields an exact sequence
\be
   K_0(R) \xrightarrow{\quad\pi\quad} K_0(R[f^{-1}]) \xrightarrow{\quad\quad} 0.
\ee

We write $P=h/f$, which is in $M_N(R[f^{-1}])$.
Then $P^2=P$, and the $R[f^{-1}]$-module 
$P (R[f^{-1}])^N$ is projective and finitely generated.
By the surjectivity of the group homomorphism $\pi$,
the $K$-theory element $[P (R[f^{-1}])^N] \in K_0(R[f^{-1}])$ has a preimage
in $K_0(R)$. Since, as mentioned, all finitely generated projective
$R$-modules are free, we thus have the following equation in $K_0(R[f^{-1}])$
\be\label{triv1}
    [P R[f^{-1}]^N] = [\pi( R^d)] = [R[f^{-1}]^d]\,.
\ee

Finally, since $f(\v k)$ is nonzero everywhere on $T^2$, $R[f^{-1}]$ is naturally embedded in 
$C(T^2)$ via the evaluation map. [Indeed, injectivity of this embedding follows 
easily from that of the similar embedding of $R$ in $C(T^2)$.]
This again yields a functor $\tilde \pi = C(T^2) \otimes_{R[f^{-1}]}$,
and a corresponding homomorphism from $K_0(R[f^{-1}])$ to $K_0(C(T^2))$.
Applying this homomorphism to \Eq{triv1} gives
\be\label{main}
    [P(\v k) C(T^2)^N]=[C(T^2)^d]\,.
\ee
The first Chern map is a map from $K_0(C(T^2))=K^0(T^2)$ to the second cohomology group
$H^2(T^2,\mathbb{Z})$. In particular, the Chern number of a given vector bundle depends only
on the $K$-theory element associated to that bundle.
\Eq{main} therefore implies that the Chern number of the bundle in question here is that
of a trivial bundle, and hence vanishes.
This is the result we desired.

We emphasize that by showing that the $K$-theory of the bundle is that of a trivial bundle,
we did not show that the bundle itself is trivial. This is unlike in the case of a non-degenerate flat band,
where not only does the vanishing of the Chern number imply the triviality of the
bundle, but our proof was based on explicit construction of a global section.
 Thus, we cannot rule out at this point that for $d>1$, flat bands arising in 
 local tight binding models can be nontrivial in other ways.
 However, absent any special symmetries, the Chern number is the topological invariant of greatest physical significance, due to
its connection with edge states and the quantized Hall effect.
Our findings imply that the Chern number always vanishes in models of the said type.

\section{Conclusion\label{conclusion}}

In this work, we have proven a theorem according to which the Chern number  for $d$-fold degenerate isolated flat bands
arising in strictly local periodic tight binding models always vanishes.
For $d=1$, we have given an elementary proof, explicitly constructing a 
nowhere vanishing global section. In the degenerate case, we have made contact
with results in $K$-theory. This proves even for $d>1$ that the residual band width
of Chern-bands defined in the literature (mostly with $d=1$) is fundamentally
necessary, so long as the underlying models are sufficiently ``simple'', i.e., have finite hopping range.
We have further presented a class of three-band models with nearest and next-nearest neighbor interactions,
whose isolated flat middle band is topologically trivial and can be tuned to an atomic limit, but has 
non-vanishing Berry curvature and Wannier states of extended support in general.
The generalization of these results to other topological invariants, in particular
the  $\mathbb{Z}_2$ invariant \cite{kanemele} of time-reversal symmetric topological insulators,
remains as an interesting problem for the future.

\ack

The work of LC, TM, and AS has been supported by NSF under Grant No. DMR-1206781.
AS would like to thank Professors R. Thomale, Jian-Xin Li, Z. Nussinov, and A. L\"auchli for insightful discussion.
Tang would like to thank Professor Max Karoubi and Yi-jun Yao for the inspiring discussions on $K$-theory. 
Tang's research is partially supported by NSF grant 0900985, and NSA grant H96230-13-1-02.

\section*{References}


\begin{thebibliography}{10}

\bibitem{Tang11}
Evelyn Tang, Jia-Wei Mei, and Xiao-Gang Wen.
\newblock {High-Temperature Fractional Quantum Hall States}.
\newblock {\em Phys. Rev. Lett.}, 106:236802, Jun 2011.

\bibitem{Sun11}
Kai Sun, Zhengcheng Gu, Hosho Katsura, and S.~Das~Sarma.
\newblock {Nearly Flatbands with Nontrivial Topology}.
\newblock {\em Phys. Rev. Lett.}, 106:236803, Jun 2011.

\bibitem{Neu11}
Titus Neupert, Luiz Santos, Claudio Chamon, and Christopher Mudry.
\newblock {Fractional Quantum Hall States at Zero Magnetic Field}.
\newblock {\em Phys. Rev. Lett.}, 106:236804, Jun 2011.

\bibitem{Sheng11}
D.~N. {Sheng}, Z.-C. {Gu}, K.~{Sun}, and L.~{Sheng}.
\newblock {Fractional quantum Hall effect in the absence of Landau levels}.
\newblock {\em Nature Communications}, 2:389, July 2011.

\bibitem{Qi11}
Xiao-Liang Qi.
\newblock Generic wave-function description of fractional quantum anomalous
  hall states and fractional topological insulators.
\newblock {\em Phys. Rev. Lett.}, 107:126803, Sep 2011.

\bibitem{Wang11}
Yi-Fei Wang, Zheng-Cheng Gu, Chang-De Gong, and D.~N. Sheng.
\newblock {Fractional Quantum Hall Effect of Hard-Core Bosons in Topological
  Flat Bands}.
\newblock {\em Phys. Rev. Lett.}, 107:146803, Sep 2011.

\bibitem{Regnault11}
N.~Regnault and B.~Andrei Bernevig.
\newblock {Fractional Chern Insulator}.
\newblock {\em Phys. Rev. X}, 1:021014, Dec 2011.

\bibitem{Ran11}
Fa~Wang and Ying Ran.
\newblock {Nearly flat band with Chern number $C=2$ on the dice lattice}.
\newblock {\em Phys. Rev. B}, 84:241103, Dec 2011.

\bibitem{Bernevig12}
B.~Andrei Bernevig and N.~Regnault.
\newblock {Emergent many-body translational symmetries of Abelian and
  non-Abelian fractionally filled topological insulators}.
\newblock {\em Phys. Rev. B}, 85:075128, Feb 2012.

\bibitem{WangPRL12}
Yi-Fei Wang, Hong Yao, Zheng-Cheng Gu, Chang-De Gong, and D.~N. Sheng.
\newblock {Non-Abelian Quantum Hall Effect in Topological Flat Bands}.
\newblock {\em Phys. Rev. Lett.}, 108:126805, Mar 2012.

\bibitem{WangPRB12}
Yi-Fei Wang, Hong Yao, Chang-De Gong, and D.~N. Sheng.
\newblock {Fractional quantum Hall effect in topological flat bands with Chern
  number two}.
\newblock {\em Phys. Rev. B}, 86:201101, Nov 2012.

\bibitem{Wu12}
Yang-Le Wu, B.~Andrei Bernevig, and N.~Regnault.
\newblock {Zoology of fractional Chern insulators}.
\newblock {\em Phys. Rev. B}, 85:075116, Feb 2012.

\bibitem{Laeuchli12}
Zhao Liu, Emil~J. Bergholtz, Heng Fan, and Andreas~M. L\"auchli.
\newblock {Fractional Chern Insulators in Topological Flat Bands with Higher
  Chern Number}.
\newblock {\em Phys. Rev. Lett.}, 109:186805, Nov 2012.

\bibitem{Bergholtz12}
Maximilian Trescher and Emil~J. Bergholtz.
\newblock {Flat bands with higher Chern number in pyrochlore slabs}.
\newblock {\em Phys. Rev. B}, 86:241111, Dec 2012.

\bibitem{Sarma12}
Shuo Yang, Zheng-Cheng Gu, Kai Sun, and S.~Das~Sarma.
\newblock {Topological flat band models with arbitrary Chern numbers}.
\newblock {\em Phys. Rev. B}, 86:241112, Dec 2012.

\bibitem{Liu13}
Tianhan Liu, C.~Repellin, B.~Andrei Bernevig, and N.~Regnault.
\newblock {Fractional Chern insulators beyond Laughlin states}.
\newblock {\em Phys. Rev. B}, 87:205136, May 2013.

\bibitem{bergholtz_review}
E.~J. {Bergholtz} and Z.~{Liu}.
\newblock {Topological Flat Band Models and Fractional Chern Insulators}.
\newblock {\em International Journal of Modern Physics B}, 27:30017, September
  2013.

\bibitem{TKNN}
D.~J. Thouless, M.~Kohmoto, M.~P. Nightingale, and M.~den Nijs.
\newblock {Quantized Hall Conductance in a Two-Dimensional Periodic Potential}.
\newblock {\em Phys. Rev. Lett.}, 49:405--408, Aug 1982.

\bibitem{haldane}
F.~D.~M. Haldane.
\newblock {Model for a Quantum Hall Effect without Landau Levels:
  Condensed-Matter Realization of the "Parity Anomaly"}.
\newblock {\em Phys. Rev. Lett.}, 61:2015--2018, Oct 1988.

\bibitem{mcgreevy}
John McGreevy, Brian Swingle, and Ky-Anh Tran.
\newblock {Wave functions for fractional Chern insulators}.
\newblock {\em Phys. Rev. B}, 85:125105, Mar 2012.

\bibitem{Qi}
Chao-Ming Jian, Zheng-Cheng Gu, and Xiao-Liang Qi.
\newblock {Momentum-space instantons and maximally localized flat-band
  topological Hamiltonians}.
\newblock {\em physica status solidi (RRL) ‚Äì Rapid Research Letters},
  7(1-2):154--156, 2013.

\bibitem{katsura}
Hosho Katsura, Isao Maruyama, Akinori Tanaka, and Hal Tasaki.
\newblock Ferromagnetism in the hubbard model with topological/non-topological
  flat bands.
\newblock {\em EPL (Europhysics Letters)}, 91(5):57007, 2010.

\bibitem{kapit}
Eliot Kapit and Erich Mueller.
\newblock Exact parent hamiltonian for the quantum hall states in a lattice.
\newblock {\em Phys. Rev. Lett.}, 105:215303, Nov 2010.

\bibitem{atakisi}
Hakan Ataki\ifmmode~\mbox{\c{s}}\else \c{s}\fi{}i and M.~\"O. Oktel.
\newblock Landau levels in lattices with long-range hopping.
\newblock {\em Phys. Rev. A}, 88:033612, Sep 2013.

\bibitem{nussinov}
Z.~{Nussinov} and J.~{van den Brink}.
\newblock {Compass and Kitaev models -- Theory and Physical Motivations}.
\newblock {\em ArXiv:1303.5922}, March 2013.

\bibitem{hastings}
M.~B. {Hastings} and T.~{Koma}.
\newblock {Spectral Gap and Exponential Decay of Correlations}.
\newblock {\em Communications in Mathematical Physics}, 265:781--804, August
  2006.

\bibitem{AlgebraI}
N.~Bourbaki.
\newblock {\em Algebra I: Chapters 1-3}.
\newblock Elements de mathematique [series]. Springer, 1998.

\bibitem{bass}
H.~Bass.
\newblock {\em {Algebraic K-theory}}.
\newblock W. A. Benjamin Inc., 1968.

\bibitem{swan}
R.~G. {Swan}.
\newblock {Projective modules over Laurent polynomial rings}.
\newblock {\em Trans. Amer. Math. Soc.}, 237:111--120, 1978.

\bibitem{quillen}
D.~{Quillen}.
\newblock {Projective modules over polynomial rings}.
\newblock {\em Invent. Math.}, 237:167--171, 1976.

\bibitem{kanemele}
C.~L. Kane and E.~J. Mele.
\newblock {${Z}_{2}$ Topological Order and the Quantum Spin Hall Effect}.
\newblock {\em Phys. Rev. Lett.}, 95:146802, Sep 2005.

\end{thebibliography}

\end{document}